\documentclass[referee]{raa}            

\usepackage{graphicx,times}             
\usepackage{natbib}
\usepackage{amssymb,amsmath}
\bibpunct{(}{)}{;}{a}{}{,}
\newcommand{\twodm}{(m-M)_0}

\usepackage{lineno}
\usepackage[pagebackref=true]{hyperref}
\bibpunct{(}{)}{;}{a}{}{,}

\def\apjl{ApJL}
\def\apjs{ApJS}
\def\aap{A\&A}

\begin{document}

\title{Identifying Distant and Highly Reddened Star Clusters: 486 Open Clusters and Candidates in the Dusty Galactic Disk
}

\volnopage{Vol.0 (20xx) No.0, 000--000}      
\setcounter{page}{1}          

\author{Zhihong He  
      \inst{1}
   \and Xiaoping Wang 
      \inst{1}
   \and Qian Cui
      \inst{1}
    \and Yangping Luo
      \inst{1}
    \and Kun Wang
      \inst{1}
   }

\institute{School of Physics and Astronomy, China West Normal University,
No.~1 Shida Road, Nanchong 637002, China;
{\it hezh@cwnu.edu.cn}\\
\vs\no {\small Received 20xx month day; accepted 20xx month day}}

\abstract{The Gaia mission has transformed the census of Galactic star clusters. Most discoveries were based on Gaia DR2 and EDR3, and the rate of new large cluster catalogues has declined since 2023. The distant and dusty part of the disk is still poorly sampled because optical data are strongly affected by extinction. We search for star clusters in the Galactic disk using Gaia DR3 astrometry and photometry, together with 2MASS near-infrared photometry. We scan trial cluster centers in Galactic position, proper motion, and parallax space, assign a Bayesian membership probability to each star, construct a relative density map, and use DBSCAN solely to separate neighboring peaks. We identify 486 clusters and candidates with astrometric quality comparable to that of previously reported clusters. Of these, 450 have no literature counterpart, while 36 are known clusters that were not recovered in earlier Gaia cluster searches. The median distance is 6.92~kpc, and 39\% of the objects lie beyond 8~kpc. The median extinction is \(A_V = 5.10\)~mag, and 16\% of the objects have \(A_V > 8\)~mag. The new objects fill a sparsely sampled region of the distance--extinction plane. Their distribution reveals the north--south displacement of the Galactic warp, and high-latitude objects are on average farther, older, and less extincted. Gaia DR3 still offers room for distant, dust-obscured clusters, and this sample provides a baseline for Gaia DR4. All cluster parameters, member star information, spatial distribution maps, and color-magnitude diagrams can be downloaded from https://zenodo.org/records/24431850 ~\citep{noc_2026_data}.
\keywords{Galaxy: stellar content --- open clusters and associations: general --- Galaxy: disk --- surveys --- methods: data analysis}
}

\authorrunning{Z.-H. He, X.-P. Wang, Q. Cui, Y.-P Luo \& K. Wang}
\titlerunning{486 Open Clusters and Candidates in the Dusty Galactic Disk}

\maketitle

\section{Introduction}\label{sec:intro}

Gaia is a billion-star surveyor that delivers astrometry at the tens of micro-arcsecond level for more than one billion sources \citep{Gaia16,gaiadr3}. Its parallaxes and proper motions have made open clusters (OCs) one of the most powerful tracers of the Galactic disk: cluster members share distance, age, and kinematics, so a single cluster provides a precise anchor in six-dimensional phase space \citep{Cantat22,Kuhn19}. The same properties make clusters useful tracers of the disk structure, including the spiral arms~\citep{Castro21} and the Galactic warp~\citep{he23warp}.

The Gaia revolution in cluster discovery started with DR2 \citep{Gaia18-Brown}. \citet{CG18} re-identified about 1\,200 known clusters and provided homogeneous memberships, while early searches found a few dozen clusters in targeted fields \citep[e.g.][]{Castro18,CG19-0,Castro19,Sim19,Liu19,Hunt21,Ferreira20,Casado21,Jaehnig21}. The application of DBSCAN~\citep{Ester96} in a five-dimensional astrometric space, together with artificial neural networks for CMD verification, raised the number of new clusters to 582 in Gaia DR2 \citep{Castro20} and to 628 in EDR3 \citep{castro22}. The DBSCAN searches of He et al.\ produced 886 nearby clusters \citep{he22b}, 1\,656 clusters at intermediate distances \citep{he23a}, and 2\,086 objects at low Galactic latitudes, including distant clusters \citep{he23b}.  The all-sky HDBSCAN catalogue of~\citet{hunt23} recovered 7\,167 clusters, of which 2\,387 are candidate new objects and 739 are new in its high-confidence subsample. Other studies employing alternative methodologies have also enriched the census~\citep[e.g.][]{hao22,li22,chi23,qin23,casado23}.

This rapid growth, however, has largely exhausted the easy parameter space. Most of the discoveries were made before the end of 2023, and the post-2023 literature is dominated by catalog cleaning, completeness estimates~\citep[e.g.][]{hunt24,Perren_2023}, and candidate confirmation rather than by the search for new star clusters. The distant and highly obscured regime remains the least explored: clusters beyond about 5~kpc are rare, and only a small fraction of the known clusters have $A_V>5$~mag. The distant survey of \citet{he23b} made a major step in this direction, but the regime is still far from complete. Near-infrared surveys have confirmed individual objects and small groups in the inner disk and at 10--14~kpc \citep{garro24}. A dedicated VVVX search found 40 new candidates in about 1100~deg$^2$ of the inner bulge and disk \citep{obasi26}, and many of these clusters are absent from Gaia-based catalogues. The Emei clusters at about 14~kpc in the outer disk \citep{he26} and the highly reddened clusters in the inner disk \citep{cui26} show that the census is still incomplete at both ends of the disk. This raises the question of whether Gaia DR3 can still find new clusters, including re-identifying known near-infrared clusters that were not recovered in previous Gaia searches.

The present work is a continuation of our series of Gaia cluster searches \citep{he22a,he22b,he23a,he23b,cui26}. We use Gaia DR3 astrometry and photometry together with 2MASS near-infrared photometry to search for clusters in the Galactic disk. The main difference from our previous works is the search strategy: instead of a blind DBSCAN search, we scan trial centers in Galactic position and proper motion, assign a Bayesian membership probability, construct a relative density map, select high-density peaks, and use DBSCAN only to separate adjacent peaks. This approach is designed for the faint, reddened, and spatially sparse regime where the local stellar density varies strongly and a fixed clustering threshold is difficult to choose.

The paper is organized as follows. Section~\ref{sec:data} describes the data and the reference catalogues. Section~\ref{sec:method} presents the search method, the visual inspection, the isochrone fitting, the classification, and the cross-match. Section~\ref{sec:results} presents the catalogue and its statistical properties. Section~\ref{sec:discussion} discusses completeness, the warp and extinction signatures, and the implications for Gaia DR3 and DR4. Section~\ref{sec:summary} summarizes our results.

\section{Data}\label{sec:data}

\subsection{Gaia DR3}

We use the Gaia DR3 five-parameter astrometric solution \citep{gaiadr3}, retaining the Galactic longitude $l$, Galactic latitude $b$, parallax $\varpi$ and its uncertainty, proper motions $\mu_{\alpha*}$ and $\mu_\delta$ and their uncertainties, the $G$-band mean magnitude, and the $G_{\rm BP}-G_{\rm RP}$ color. We restrict the search to the Galactic disk, $|b|\leq 15^\circ$, where the cluster population is concentrated and where the dust column is largest. No hard magnitude cut is applied beyond the requirement that the astrometric solution and the two photometric bands are available.

\subsection{2MASS near-infrared photometry}

Optical photometry alone is severely limited in the dusty disk: at $A_V\gtrsim5$~mag, the main sequence of a distant cluster is shifted to red colors and becomes faint in $G$, and the CMD is easily confused with the field. We therefore match the Gaia sources with the 2MASS point-source catalogue \citep{Skrutskie06} using a position tolerance of 1~arcsec, and retain the $J$ and $K_s$ magnitudes. The near-infrared bands are used as an independent reference for the isochrone fitting, not as an additional discovery channel. About 69\% of the member stars in our final catalogue have a valid $J$-band counterpart.

\subsection{Reference catalogues}

The search is performed after removing the members of known clusters, and the final objects are cross-matched with the literature. The reference material includes Gaia-based catalogues \citep{CG18,CG20,Castro18,Castro19,Castro20,castro22,hao22,he21,he22a,he22b,he23a,he23b,hunt23,chi23,li22} and the Unified Cluster Catalogue \citep{Perren_2023}, as well as pre-Gaia catalogues that do not provide Gaia astrometry, such as the Milky Way Star Clusters catalogue \citep{Kharchenko13}, the Dias catalogue \citep{Dias02,dias21}, and the Bica catalogue \citep{Bica19}. The two families are used differently: the Gaia-based catalogues allow a full astrometric match, while the pre-Gaia catalogues mainly provide a positional check and a literature name. Section~\ref{sec:crossmatch} describes the matching criterion.

\section{Method}\label{sec:method}

The method is designed to find sparse, distant clusters whose members occupy a small volume in proper-motion space but are not necessarily the densest group in the field. It has five steps: (i) trial centers on a grid in Galactic longitude and latitude with a step of $0.01^\circ$, in proper motion with a step of $0.5$~mas~yr$^{-1}$, and in parallax from 0.05 to 0.50~mas with a step of 0.05~mas; (ii) a Bayesian membership probability for each star; (iii) a relative density map in the Galactic plane; (iv) peak selection and separation of adjacent peaks; and (v) visual inspection and isochrone fitting. Similar density-based searches have been applied to near-infrared data \citep{obasi26}. Here, however, DBSCAN \citep{Ester96} is not used as a blind cluster finder. It is applied only to separate distinct high-density peaks after the peak selection.

\subsection{Trial centers and search radius}

For a trial center $(l_0,b_0,\mu_{c,\alpha}^*,\mu_{c,\delta})$ and a trial parallax $\varpi_c$, we select the stars in a circular region of radius
\begin{equation}
r_0 = \max\left(100\,{\rm pc}\times\frac{57.3}{d}, \; 0.3^\circ\right),
\label{eq:search_radius_new}
\end{equation}
where $d=1000/\varpi_c$ is the distance in pc. The radius is clamped between $0.3^\circ$ and $5^\circ$.

We estimate the field parameters using the median and the median absolute deviation (MAD):
\begin{equation}
\varpi_f = \mathrm{median}(\varpi_{\rm field}), \qquad
\sigma_{\varpi,f} = 1.4826\times\mathrm{MAD}(\varpi_{\rm field}).
\label{eq:field_plx}
\end{equation}
The same procedure is applied to the two proper-motion components, giving $\mu_{f,\alpha}^*$, $\mu_{f,\delta}$, $\sigma_{\mu,\alpha,f}$, and $\sigma_{\mu,\delta,f}$.

\subsection{Intrinsic cluster kinematics}

The intrinsic parallax dispersion of cluster members is modelled as
\begin{equation}
\sigma_{\varpi,c} = 0.03~\mathrm{mas},
\label{eq:plx_intrinsic}
\end{equation}
and the intrinsic proper-motion dispersion is
\begin{equation}
\sigma_{\mu,c}^2 = 0.04 + \left(\frac{\varpi_c}{4.74}\right)^2,
\label{eq:pm_intrinsic}
\end{equation}
where $\varpi_c$ is in mas and the result is in mas~yr$^{-1}$. We use the same intrinsic dispersion for the two proper-motion components. We adopt a prior membership probability $\phi=0.02$ for the Galactic plane.

\subsection{Bayesian membership probability}

For each star, we compute the likelihoods under the cluster ($H_c$) and field ($H_f$) hypotheses. The parallax likelihood is
\begin{equation}
\mathcal{L}(\varpi|H_c) = \frac{1}{\sqrt{2\pi\sigma_{\varpi,c,{\rm tot}}^2}}\exp\left[-\frac{(\varpi-\varpi_c)^2}{2\sigma_{\varpi,c,{\rm tot}}^2}\right],
\label{eq:L_plx_c}
\end{equation}
with
\begin{equation}
\sigma_{\varpi,c,{\rm tot}}^2 = \sigma_{\varpi,{\rm meas}}^2 + \sigma_{\varpi,c}^2,
\label{eq:plx_var_tot}
\end{equation}
and for the field,
\begin{equation}
\mathcal{L}(\varpi|H_f) = \frac{1}{\sqrt{2\pi\sigma_{\varpi,f,{\rm tot}}^2}}\exp\left[-\frac{(\varpi-\varpi_f)^2}{2\sigma_{\varpi,f,{\rm tot}}^2}\right],
\label{eq:L_plx_f}
\end{equation}
with $\sigma_{\varpi,f,{\rm tot}}^2=\sigma_{\varpi,{\rm meas}}^2+\sigma_{\varpi,f}^2$. Here $\sigma_{\varpi,{\rm meas}}$ is the formal parallax uncertainty from Gaia. The proper-motion likelihoods are products of two Gaussians,
\begin{equation}
\mathcal{L}(\mu_\alpha^*,\mu_\delta|H_c) = \frac{1}{2\pi\sigma_{\mu,\alpha,c,{\rm tot}}\sigma_{\mu,\delta,c,{\rm tot}}}
\exp\left[-\frac{(\mu_\alpha^*-\mu_{c,\alpha}^*)^2}{2\sigma_{\mu,\alpha,c,{\rm tot}}^2}
-\frac{(\mu_\delta-\mu_{c,\delta})^2}{2\sigma_{\mu,\delta,c,{\rm tot}}^2}\right],
\label{eq:L_pm_c}
\end{equation}
\begin{equation}
\mathcal{L}(\mu_\alpha^*,\mu_\delta|H_f) = \frac{1}{2\pi\sigma_{\mu,\alpha,f,{\rm tot}}\sigma_{\mu,\delta,f,{\rm tot}}}
\exp\left[-\frac{(\mu_\alpha^*-\mu_{f,\alpha}^*)^2}{2\sigma_{\mu,\alpha,f,{\rm tot}}^2}
-\frac{(\mu_\delta-\mu_{f,\delta})^2}{2\sigma_{\mu,\delta,f,{\rm tot}}^2}\right],
\label{eq:L_pm_f}
\end{equation}
with $\sigma_{\mu,\alpha,c,{\rm tot}}^2=\sigma_{\mu,{\rm meas}}^2+\sigma_{\mu,c}^2$ and $\sigma_{\mu,\alpha,f,{\rm tot}}^2=\sigma_{\mu,{\rm meas}}^2+\sigma_{\mu,\alpha,f}^2$, and likewise for $\delta$. Here $\sigma_{\mu,{\rm meas}}$ is the formal proper-motion uncertainty from Gaia. The joint likelihoods are
\begin{equation}
\mathcal{L}(\varpi,\mu_\alpha^*,\mu_\delta|H_c) = \mathcal{L}(\varpi|H_c)\,\mathcal{L}(\mu_\alpha^*,\mu_\delta|H_c),
\label{eq:L_joint_c}
\end{equation}
\begin{equation}
\mathcal{L}(\varpi,\mu_\alpha^*,\mu_\delta|H_f) = \mathcal{L}(\varpi|H_f)\,\mathcal{L}(\mu_\alpha^*,\mu_\delta|H_f).
\label{eq:L_joint_f}
\end{equation}
Applying Bayes' theorem, the membership probability is
\begin{equation}
p_c = \frac{\phi\mathcal{L}({\rm data}|H_c)}{\phi\mathcal{L}({\rm data}|H_c) + (1-\phi)\mathcal{L}({\rm data}|H_f)}.
\label{eq:posterior}
\end{equation}
We retain stars with $p_c>0.05$ for the density analysis.

\subsection{Relative density map and peaks}

We perform a two-dimensional Gaussian kernel density estimate \citep{Parzen1962} on the projected plane:
\begin{equation}
\rho_c(x,y) = \sum_{i=1}^{N} p_{c,i}\,K_h(x-x_i,y-y_i),
\label{eq:kde}
\end{equation}
with
\begin{equation}
K_h(dx,dy) = \frac{1}{2\pi h^2}\exp\left(-\frac{dx^2+dy^2}{2h^2}\right),
\label{eq:kernel}
\end{equation}
and a fixed bandwidth $h=0.01^\circ$. We then subtract the median density of the region and divide by its standard deviation,
\begin{equation}
\delta = \frac{\rho_c - \mathrm{median}(\rho_c)}{\mathrm{std}(\rho_c)},
\label{eq:relative_density}
\end{equation}
and select local maxima with $\delta>5$ as candidate centers. Here $\delta$ is already in units of the standard deviation of the density field, $\sigma=\mathrm{std}(\rho_c)$; the $5\sigma$ cut is a conservative choice and is not a fit parameter. DBSCAN is then applied only to separate adjacent high-density peaks; its parameters are not part of the selection. The member boundary of each candidate is defined by the contour $\rho_c=\mathrm{median}(\rho_c)+2\sigma$, where $\sigma=\mathrm{std}(\rho_c)$.

\subsection{Visual inspection and isochrone fitting}\label{sec:fit}

Each candidate is inspected by eye in the CMD. We keep candidates with a clear main sequence, a plausible turn-off, and members that are not dominated by an unrelated sequence. A candidate with a clear but contaminated CMD is kept and marked as lower quality (Section~\ref{sec:class}). A candidate with no coherent sequence is discarded.

The retained candidates are fitted with PARSEC isochrones \citep{Bressan12}, using the Gaia $G$, $G_{\rm BP}$, $G_{\rm RP}$ and the 2MASS $J$, $K_s$ photometry. The metallicity is fixed at solar ($\rm [M/H]=0$). The extinction coefficient is computed with the polynomial function provided in the Gaia DR2 auxiliary data \citep{Danielski18,gaia_cmd}. This function gives the extinction in the Gaia and 2MASS bands as a function of color and $A_V$. The fit is performed manually: we start from a distance modulus derived from the error-weighted mean parallax of the member stars, after adding a zero-point offset of 0.02~mas, and we scan a narrow range corresponding to $\pm0.05$~mas in parallax. Within this range, the distance modulus, extinction, and age are adjusted until the isochrone reproduces the observed sequence. The adopted parameters are the distance modulus $\twodm$, the extinction $A_V$ at 5500~\AA\ (approximately $A_V$), and $\log(\rm age/yr)$.

\subsection{Classification}\label{sec:class}

The candidates are divided into three classes according to their central concentration and the quality of their CMDs:
\begin{itemize}
\item class~1: a concentrated center and a clear main sequence, even if the CMD has some contamination, or a clear evolutionary sequence;
\item class~2: a center less concentrated than class~1 but still relatively concentrated, or a CMD with an unclear isochrone while the concentration is acceptable;
\item class~3: a less concentrated center and a contaminated CMD, often with a short or partly disrupted main sequence, so that the main sequence is not clear.
\end{itemize}
The classes are assigned by visual inspection and are intended as a quality flag, not as a physical classification. In particular, class~3 does not mean a higher extinction; it denotes a looser and less secure object.

\subsection{Cross-match and naming}\label{sec:crossmatch}

The cross-match is performed in two stages. First, we compare each candidate with the Gaia-based catalogues using the member-star astrometry. A literature cluster is a counterpart when its position, parallax, and proper motion agree with the median values of the members within $1\sigma$. For matches between $1\sigma$ and $3\sigma$, we check the CMD and the member distribution by eye to confirm that the object is the same cluster. Second, we compare the remaining candidates with the pre-Gaia catalogues, which do not provide Gaia astrometry, using a positional criterion: if a literature object falls inside $r_{90}$, the angular radius containing 90\% of the members, the candidate is considered known and retains the literature name. If no counterpart is found in either stage, the candidate is considered new and is assigned a CWNU identifier. The positional criterion is deliberately generous, so that known clusters are not re-numbered; the price is that a small number of genuinely new objects may be absorbed into a nearby literature entry.

\section{Results}\label{sec:results}

\subsection{The catalogue}

The search yields 486 clusters and candidates. Of these, 450 have no literature counterpart within $r_{90}$ and are new\footnote{Among these, Barcelos~2 was found independently by us and by Bruno Alessi; through communication it was designated Barcelos~2.  Another cluster, Qingcheng, was identified by us previously but has not yet been published and is currently under new observations.  All the remaining new clusters have been assigned ``CWNU'' designations.}; the remaining 36 retain literature names. The sample contains 264 class~1, 194 class~2, and 28 class~3 objects, and 24\,749 member stars. The median number of members per object is 46, with a range from 16 to 206. Table~1 gives a sample of the cluster parameters; the full table is provided as online material. Table~2 gives a sample of the member-star catalogue.

\begin{table*}[htbp]
\centering
\caption{A sample of the cluster catalogue. The full table contains 486 objects and is available as online material. $l$ and $b$ are in degrees, $r_{50}$ and $r_{90}$ in degrees, $\varpi$ in mas, proper motions in $\mathrm{mas\,yr^{-1}}$, $A_V$ in mag, and $(m-M)_0$ in mag. The class is 1, 2, or 3 (Section~\ref{sec:class}).}
\label{tab:params}
\scriptsize
\setlength{\tabcolsep}{3pt}
\renewcommand{\arraystretch}{1.15}
\resizebox{\textwidth}{!}{%
\begin{tabular}{lccccccccccccccc}
\hline
Name & Class & $l$ & $b$ & $r_{50}$ & $r_{90}$ & $N$ & $\varpi$ & $\sigma_\varpi$ & $\mu_{\alpha*}$ & $\sigma_{\mu\alpha}$ & $\mu_\delta$ & $\sigma_{\mu\delta}$ & $\log\,\mathrm{age}$ & $A_V$ & $(m-M)_0$ \\
 & & [deg] & [deg] & [deg] & [deg] & & [mas] & [mas] & [$\mathrm{mas\,yr^{-1}}$] & [$\mathrm{mas\,yr^{-1}}$] & [$\mathrm{mas\,yr^{-1}}$] & [$\mathrm{mas\,yr^{-1}}$] & [dex] & [mag] & [mag] \\
\hline
ASCC\_46 & 1 & 264.263 & -7.238 & 0.030 & 0.042 & 43 & 0.05 & 0.17 & -1.21 & 0.24 & 2.01 & 0.19 & 6.90 & 1.9 & 16.0 \\
CWNU\_783 & 1 & 1.262 & 0.441 & 0.014 & 0.031 & 25 & 0.03 & 0.25 & -1.21 & 0.27 & -3.13 & 0.15 & 7.50 & 12.9 & 14.3 \\
Alessi\_J1616.2-5136 & 1 & 331.867 & -0.624 & 0.015 & 0.027 & 64 & 0.18 & 0.11 & -3.64 & 0.20 & -3.78 & 0.13 & 6.20 & 6.9 & 14.1 \\
CWNU\_598 & 1 & 272.761 & -1.760 & 0.023 & 0.039 & 45 & -0.01 & 0.15 & -4.02 & 0.24 & 4.17 & 0.33 & 8.10 & 6.1 & 16.4 \\
CWNU\_791 & 2 & 86.216 & 0.932 & 0.015 & 0.028 & 56 & 0.17 & 0.15 & -3.02 & 0.23 & -4.11 & 0.23 & 8.25 & 5.9 & 14.2 \\
CWNU\_964 & 2 & 17.700 & -0.218 & 0.025 & 0.045 & 43 & -0.01 & 0.20 & -2.35 & 0.29 & -4.17 & 0.27 & 8.15 & 8.3 & 14.6 \\
DBSB\_111 & 2 & 344.967 & 1.794 & 0.026 & 0.044 & 48 & 0.56 & 0.14 & -0.13 & 0.24 & -1.08 & 0.21 & 6.65 & 4.6 & 11.4 \\
CWNU\_965 & 2 & 306.772 & 0.050 & 0.024 & 0.038 & 90 & 0.22 & 0.15 & -5.27 & 0.26 & -1.23 & 0.20 & 8.10 & 4.3 & 13.6 \\
CWNU\_796 & 3 & 66.410 & -0.353 & 0.026 & 0.045 & 76 & 0.17 & 0.20 & -2.45 & 0.26 & -4.05 & 0.21 & 8.80 & 7.2 & 13.1 \\
CWNU\_797 & 3 & 5.973 & -2.743 & 0.045 & 0.072 & 47 & 0.12 & 0.13 & 0.61 & 0.34 & -4.92 & 0.28 & 9.30 & 3.6 & 14.0 \\
CWNU\_798 & 3 & 340.782 & 0.037 & 0.031 & 0.050 & 72 & 0.29 & 0.13 & -2.21 & 0.27 & -2.94 & 0.25 & 8.60 & 7.1 & 12.8 \\
CWNU\_981 & 3 & 329.132 & 0.415 & 0.049 & 0.068 & 61 & 0.27 & 0.16 & -1.79 & 0.26 & -3.03 & 0.19 & 8.15 & 8.0 & 13.1 \\
\hline
\end{tabular}%
}
\end{table*}

\begin{table*}[htbp]
\centering
\caption{A sample of the member-star catalogue for CWNU~566. The full table contains 24\,749 stars and is available as online material.}
\label{tab:members}
\scriptsize
\setlength{\tabcolsep}{5pt}
\renewcommand{\arraystretch}{1.15}
\resizebox{\textwidth}{!}{%
\begin{tabular}{lcccccccccc}
\hline
Name & $l$ & $b$ & $\varpi$ & $\sigma_\varpi$ & $\mu_{\alpha*}$ & $\sigma_{\mu\alpha}$ & $\mu_\delta$ & $\sigma_{\mu\delta}$ & $G$ & $G_{\rm BP}-G_{\rm RP}$ \\
 & [deg] & [deg] & [mas] & [mas] & [$\mathrm{mas\,yr^{-1}}$] & [$\mathrm{mas\,yr^{-1}}$] & [$\mathrm{mas\,yr^{-1}}$] & [$\mathrm{mas\,yr^{-1}}$] & [mag] & [mag] \\
\hline
CWNU\_566 & 156.591 & -6.360 & 0.25 & 0.14 & -0.30 & 0.16 & 0.25 & 0.10 & 18.08 & 1.26 \\
CWNU\_566 & 156.589 & -6.334 & 0.47 & 0.23 & -0.45 & 0.24 & 0.08 & 0.16 & 18.78 & 1.17 \\
CWNU\_566 & 156.579 & -6.341 & 0.02 & 0.06 & 0.16 & 0.06 & -0.33 & 0.04 & 16.44 & 1.70 \\
CWNU\_566 & 156.576 & -6.324 & 0.43 & 0.22 & 0.51 & 0.24 & -0.29 & 0.15 & 18.69 & 1.17 \\
CWNU\_566 & 156.572 & -6.361 & 0.23 & 0.43 & -0.15 & 0.47 & 0.21 & 0.29 & 19.50 & 1.20 \\
CWNU\_566 & 156.562 & -6.341 & 0.24 & 0.10 & -0.02 & 0.10 & 0.10 & 0.07 & 17.46 & 0.59 \\
CWNU\_566 & 156.566 & -6.316 & 0.24 & 0.14 & 0.11 & 0.16 & -0.15 & 0.10 & 18.07 & 0.61 \\
CWNU\_566 & 156.565 & -6.314 & 0.07 & 0.19 & 0.49 & 0.22 & -0.06 & 0.13 & 18.49 & 1.27 \\
\hline
\end{tabular}%
}
\end{table*}

\subsection{Spatial distribution}

The 486 objects are not distributed uniformly across the disk. The sample covers the full longitude range, but it is concentrated toward the inner Galaxy: 73\% lie at $l<90^\circ$ or $l>270^\circ$, and 74\% lie within $2^\circ$ of the plane. This concentration is expected from the combination of the disk surface density and the dust column, which makes the inner disk both richer in clusters and harder to search.

\begin{figure*}
\begin{center}
\includegraphics[width=1.0\linewidth]{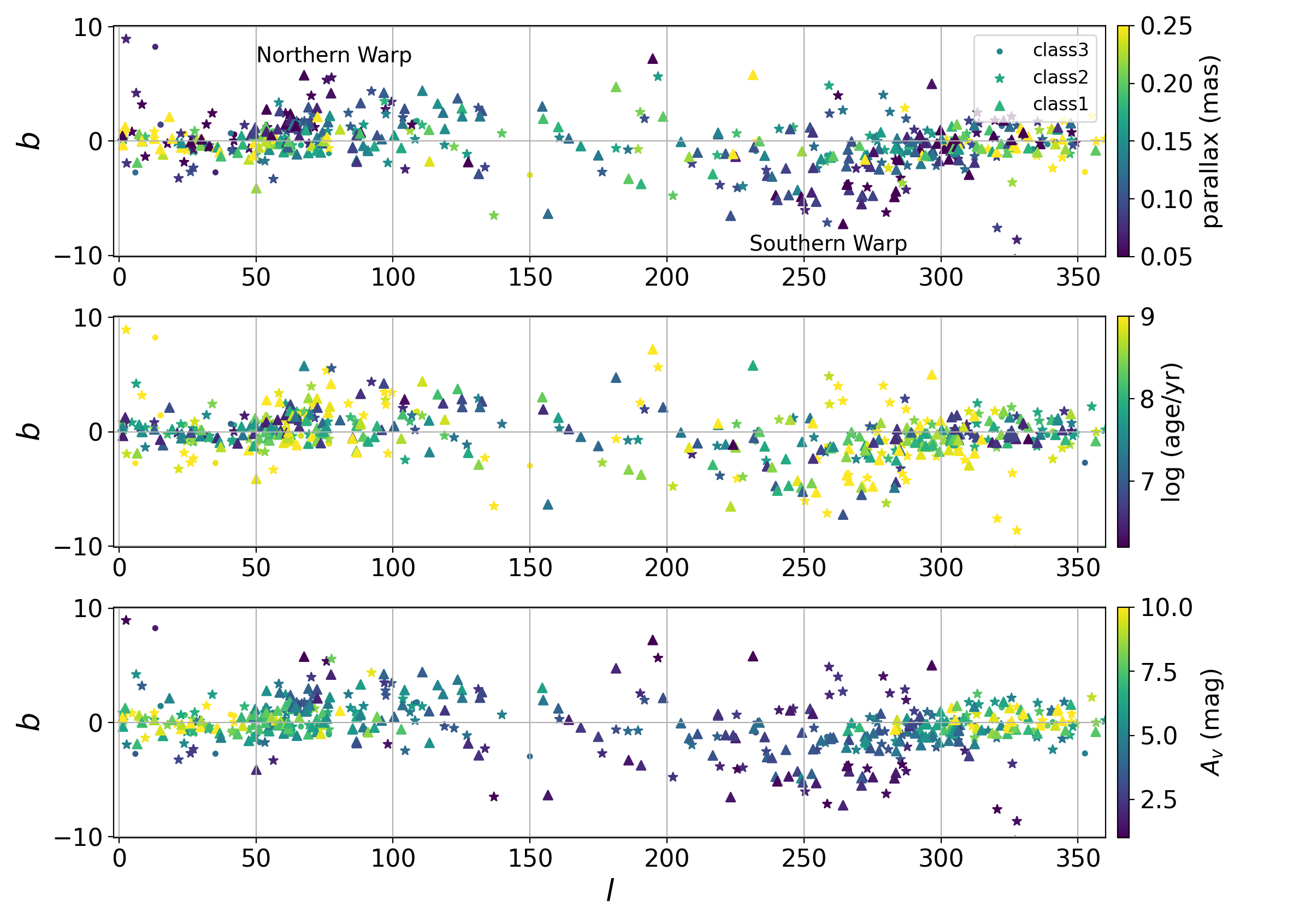}
\caption{Galactic distribution of the 486 clusters and candidates. From top to bottom, the points are color-coded by parallax, $\log(\rm age/yr)$, and $A_V$. The two high-latitude groups are marked.}
\label{fig:lb}
\end{center}
\end{figure*}

The latitude distribution is narrow near the plane and broadens at high latitude. Two groups are conspicuous. A northern group at $l\approx60^\circ$--$110^\circ$ extends to $b\approx+6^\circ$, and a southern group at $l\approx240^\circ$--$290^\circ$ extends to $b\approx-7.5^\circ$. The color coding in Figure~\ref{fig:lb} shows that the high-latitude groups are also among the most distant and least extincted objects, while the inner-Galaxy objects have higher extinction. The median $A_V$ is 6.0~mag at $l<90^\circ$ or $l>270^\circ$, compared with 3.3~mag outside this range. The distinction is not sharp, and the three classes are mixed across the disk.

\subsection{Parallax, extinction, and age distributions}

The parameter distributions are broad and asymmetric, as expected for a sample selected over a wide range of distance and extinction. The parallax distribution peaks at 0.1--0.2~mas and has a tail to about 0.8~mas; the median is 0.13~mas. Twenty-five objects have non-positive parallaxes. We keep them because their photometry still constrains the cluster sequence, and their distances are taken from the isochrone fit rather than from the inverse parallax.

The extinction distribution peaks at 4--6~mag and extends to 13.6~mag. The median is 5.10~mag; 16\% of the objects have $A_V>8$~mag, and 35 have $A_V>10$~mag. It is the signature of sightlines through the dustiest inner-disk fields, where optical photometry alone would lose the main sequence.

\begin{figure*}
\begin{center}
\includegraphics[width=1.0\linewidth]{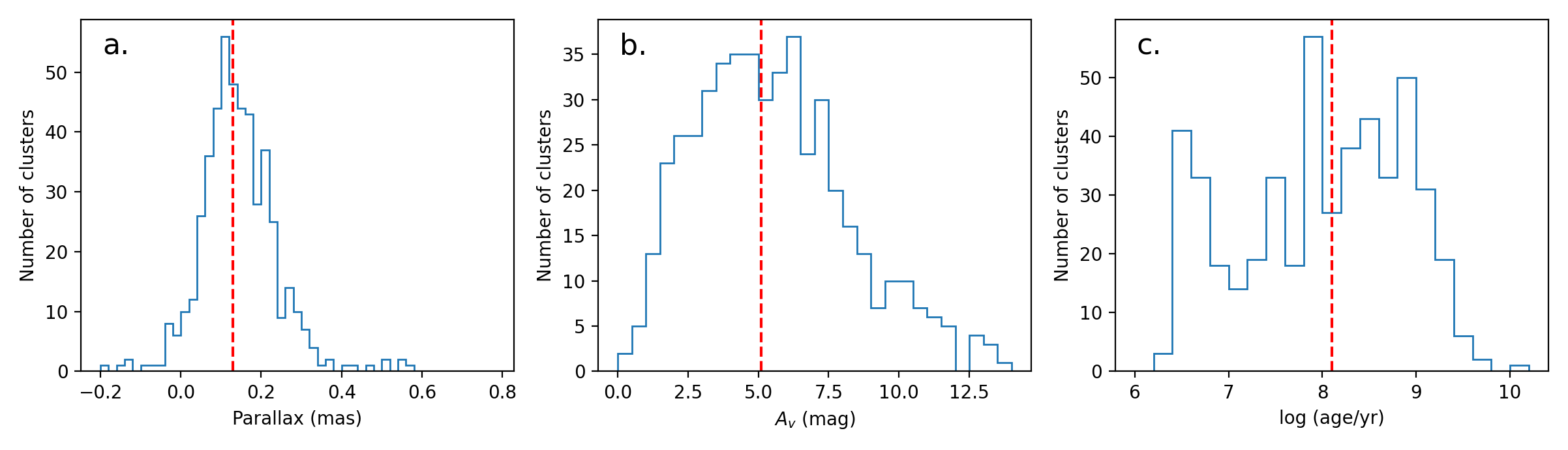}
\caption{Distributions of parallax (a), extinction (b), logarithmic age (c), and member number (d) for the 486 objects. The red dashed line marks the median.}
\label{fig:hist}
\end{center}
\end{figure*}

The age distribution ranges from $\log(\rm age/yr)=6.2$ to 10.1, with a median of 8.10; 59 objects are older than 1~Gyr. The member-number distribution peaks at 30--60, with a median of 46 and a range from 16 to 206. The sample therefore contains both sparse and rich objects, and both young and old clusters. Figure~\ref{fig:hist} summarizes these four distributions.

\begin{figure*}
\begin{center}
\includegraphics[width=0.7\linewidth]{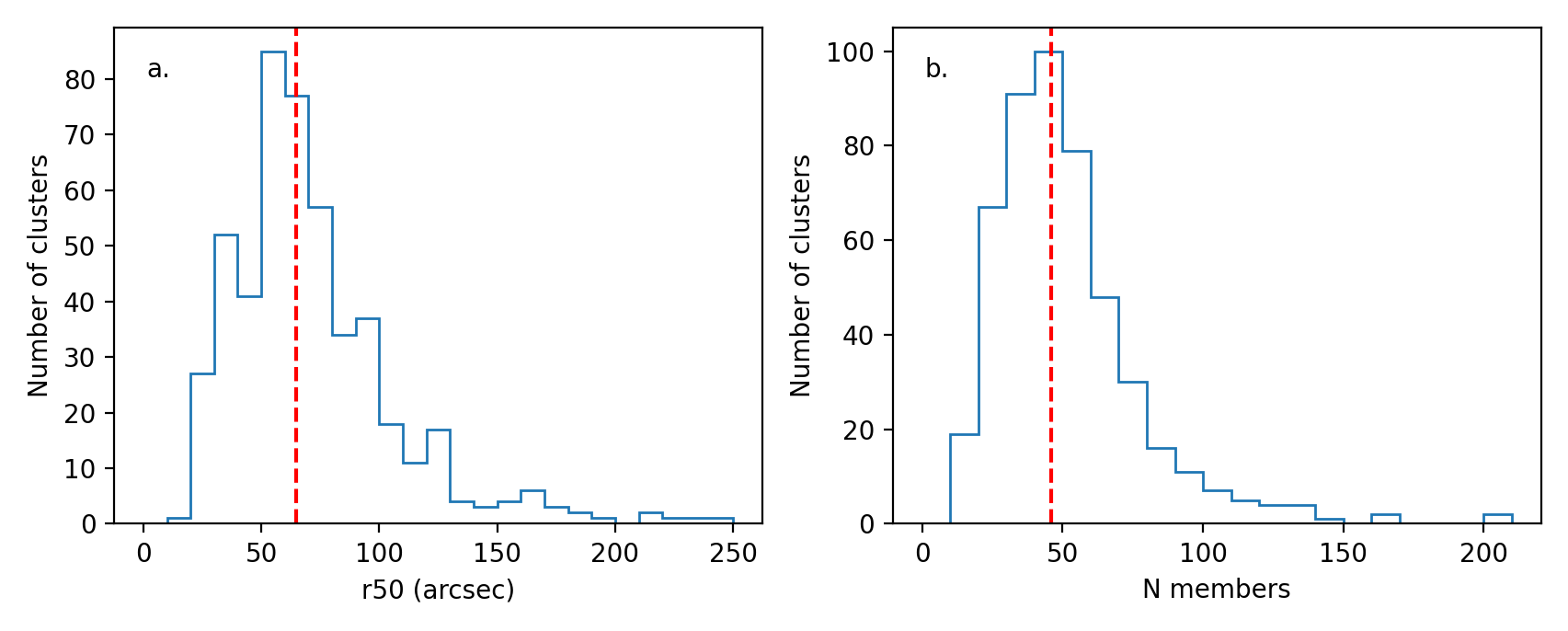}
\caption{Distributions of $r_{50}$ (a) and member number (b) for the 486 objects. The red dashed line marks the median.}
\label{fig:hist2}
\end{center}
\end{figure*}

The sizes and richnesses are correlated with the quality class. The median $r_{50}$ is 64.8~arcsec, and the median member number is 46. Class~1 has a median $r_{50}$ of 56.6~arcsec and 42 members, class~2 has 71.2~arcsec and 49 members, and class~3 has 142.2~arcsec and 86 members. The larger radii and member numbers of class~3 agree with the looser appearance of these objects. Their median extinction is 5.1~mag, which is not higher than that of class~1 (5.6~mag). Class~3 is therefore a quality class, not an extinction class. Figure~\ref{fig:hist2} shows the two distributions.

\subsection{Examples}

A useful test of the catalogue is whether it recovers objects that are known from near-infrared surveys but are difficult in the optical. Stephenson~2 and FSR~1703 are two such clusters. Both are heavily reddened, with $A_V=13.1$ and 11.9~mag, and both were found in near-infrared surveys \citep{Stephenson90,Froebrich07}. Their Gaia CMDs and 2MASS CMDs show clear sequences, and the fit gives plausible ages and distances. The recovery of these two objects shows that Gaia DR3 still adds members and parameters to known obscured clusters.

\begin{figure*}
\begin{center}
\includegraphics[width=1.0\linewidth]{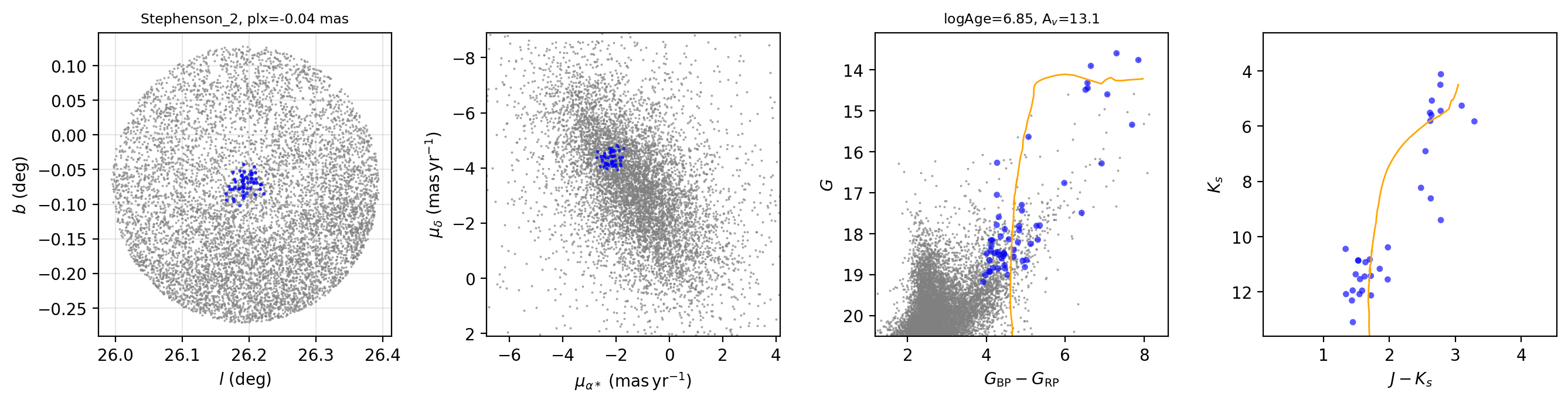}
\includegraphics[width=1.0\linewidth]{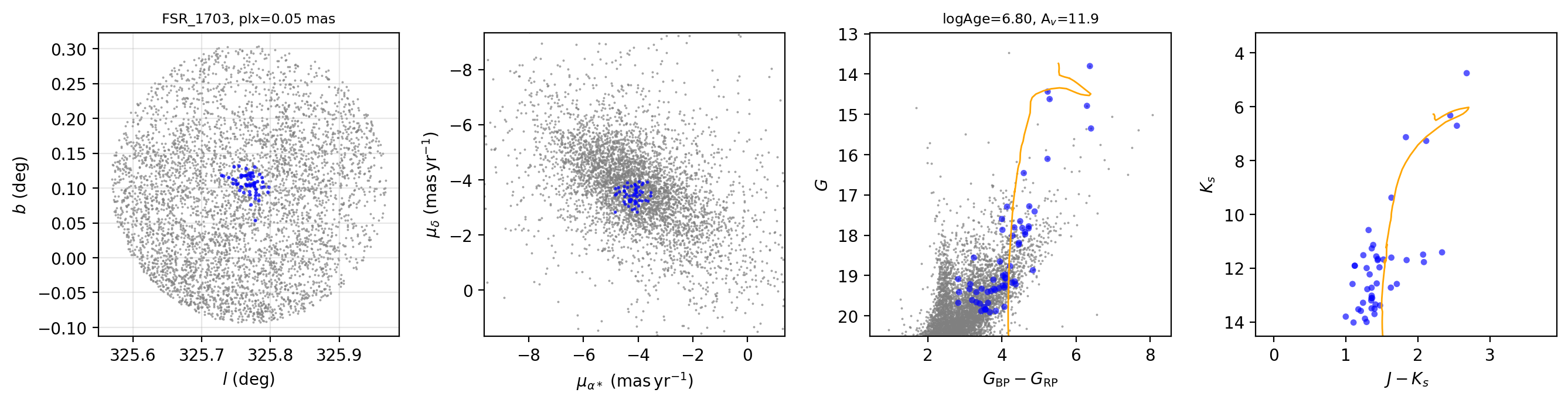}
\caption{Stephenson~2 (top) and FSR~1703 (bottom): Galactic distribution, proper motion, Gaia CMD, and 2MASS CMD. The blue points are members; the gray background points represent field stars at the same parallax.}
\label{fig:ir}
\end{center}
\end{figure*}

\begin{figure*}
\begin{center}
\includegraphics[width=1.0\textwidth]{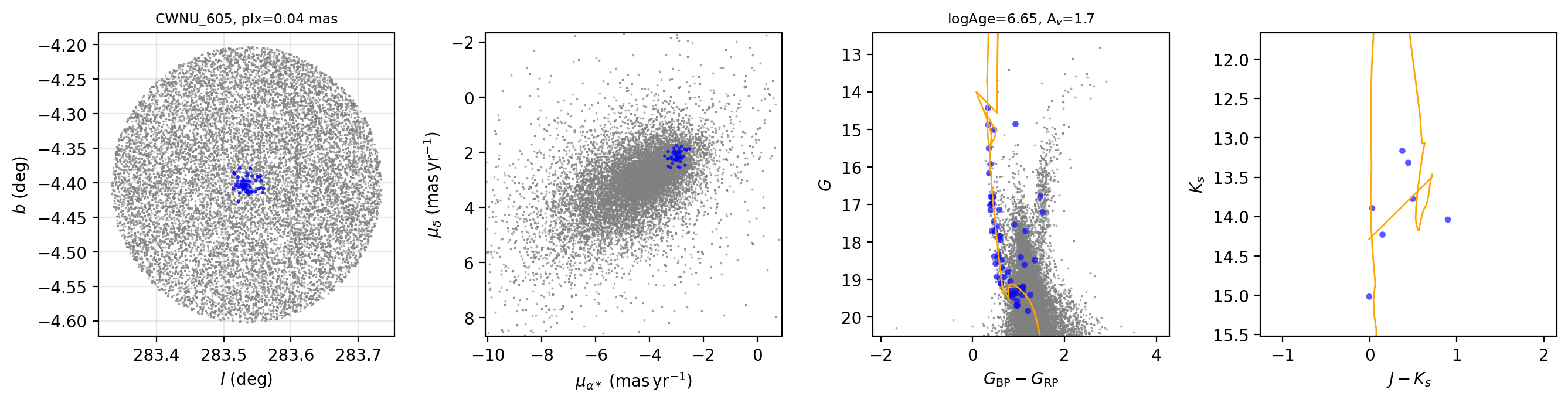}
\includegraphics[width=1.0\textwidth]{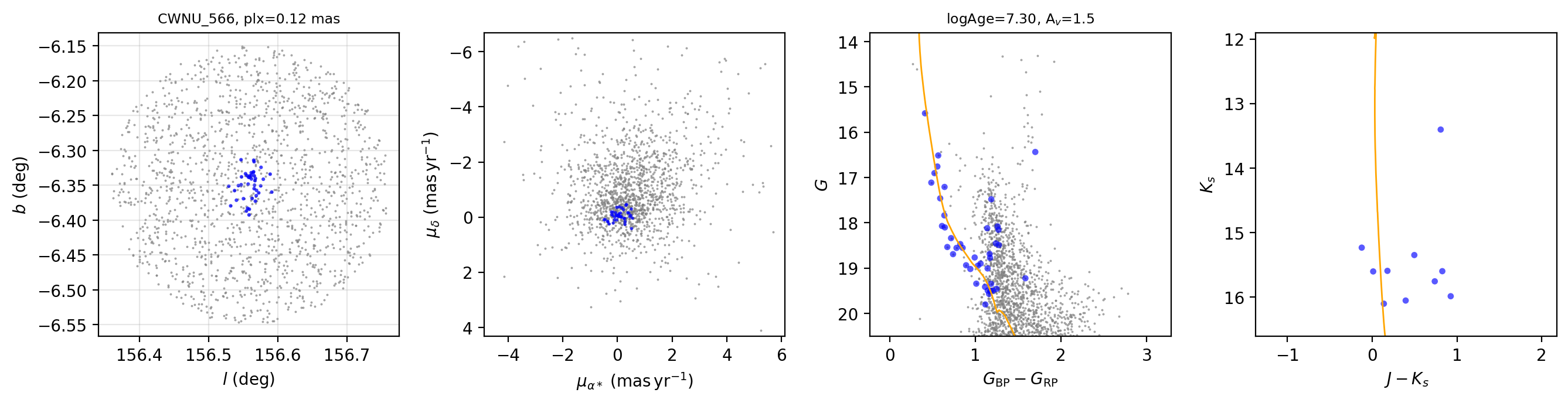}
\includegraphics[width=1.0\textwidth]{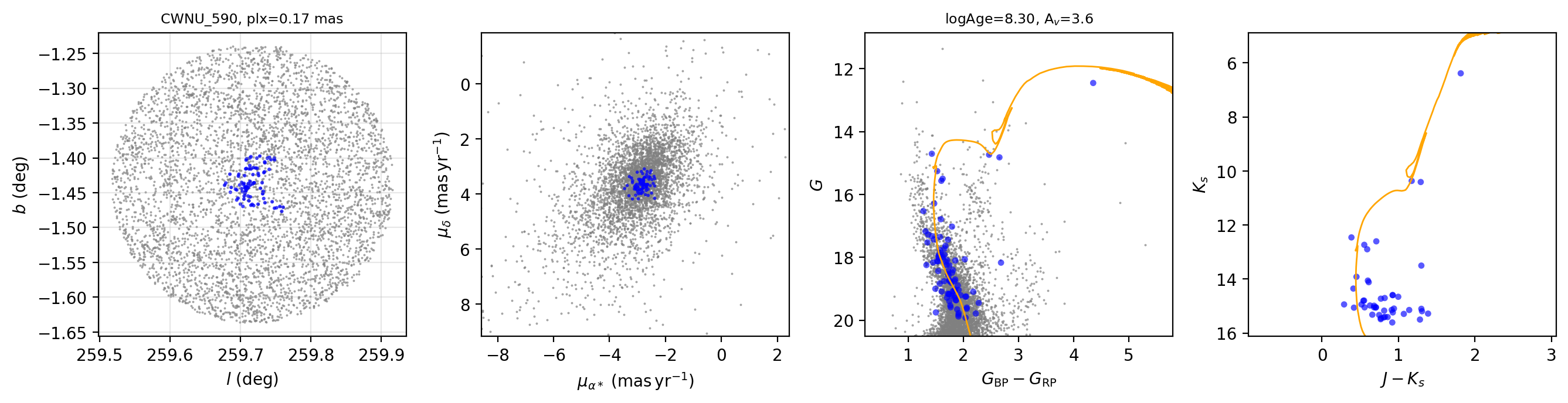}
\includegraphics[width=1.0\textwidth]{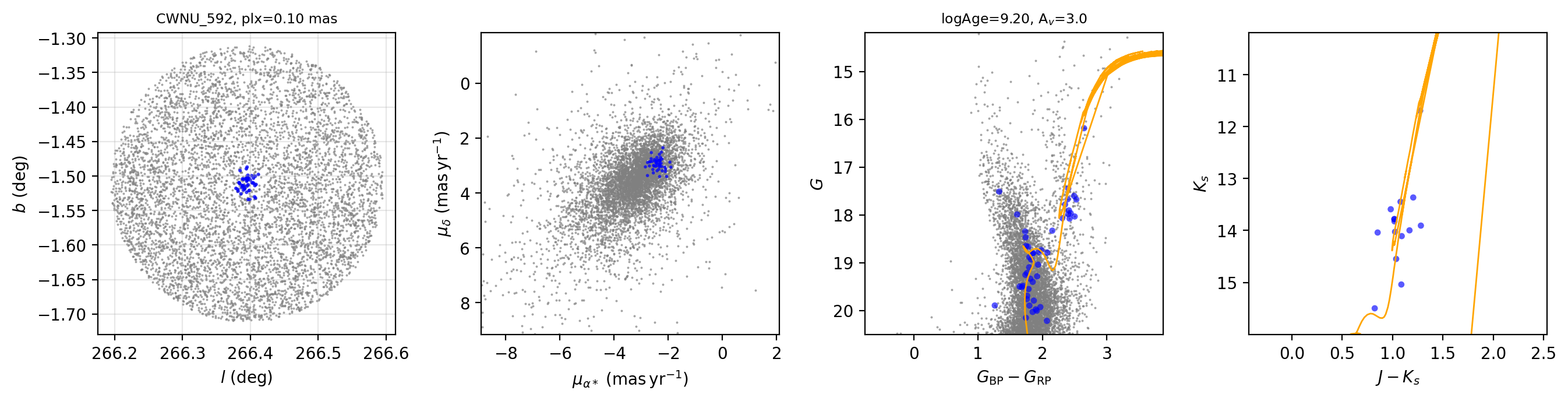}
\caption{Four class~1 clusters with increasing age: CWNU~605 ($\log$ age 6.6), CWNU~566 (7.3), CWNU~590 (8.3), and CWNU~592 (9.2). Each row shows the Galactic distribution, proper motion, Gaia CMD, and 2MASS CMD.}
\label{fig:class1}
\end{center}
\end{figure*}

The class~1 examples (Figure~\ref{fig:class1} ) illustrate the range of cluster ages. CWNU~605 ($\log$ age 6.6), CWNU~566 (7.3), CWNU~590 (8.3), and CWNU~592 (9.2) form a sequence from young to old. In all four the main sequence is well defined, and the 2MASS CMD follows the same sequence out to the redder members. We note that, at low extinction, fewer 2MASS member stars are recovered by the cross-match, so that the 2MASS photometry becomes less useful; in this regime, however, the Gaia CMD alone is sufficient to fit the isochrones. The turn-off moves as expected with age, which gives confidence in the fitted ages. 

\begin{figure*}
\begin{center}
\includegraphics[width=\textwidth]{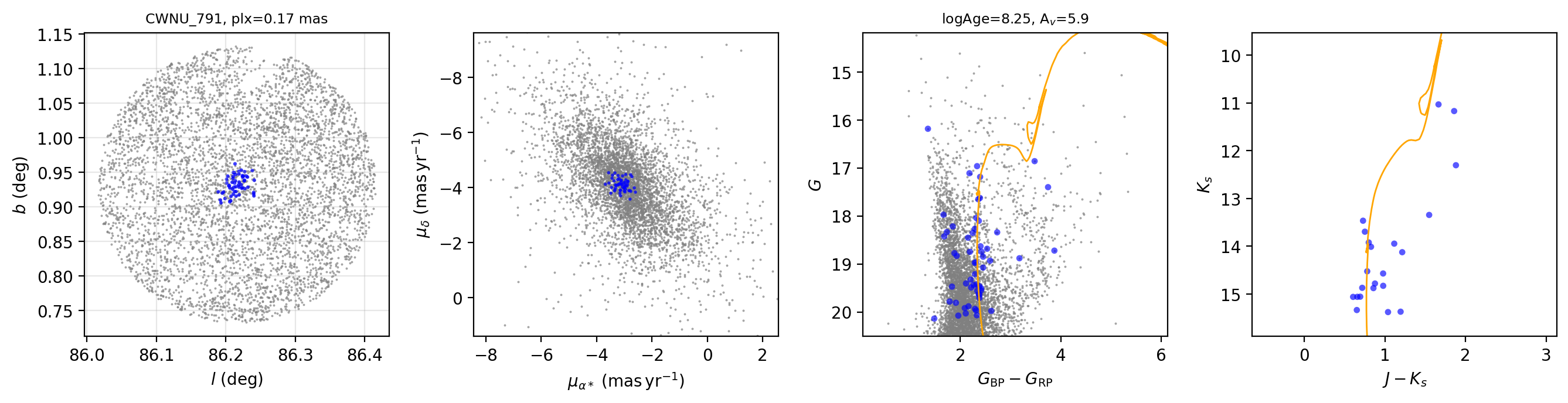}
\includegraphics[width=\textwidth]{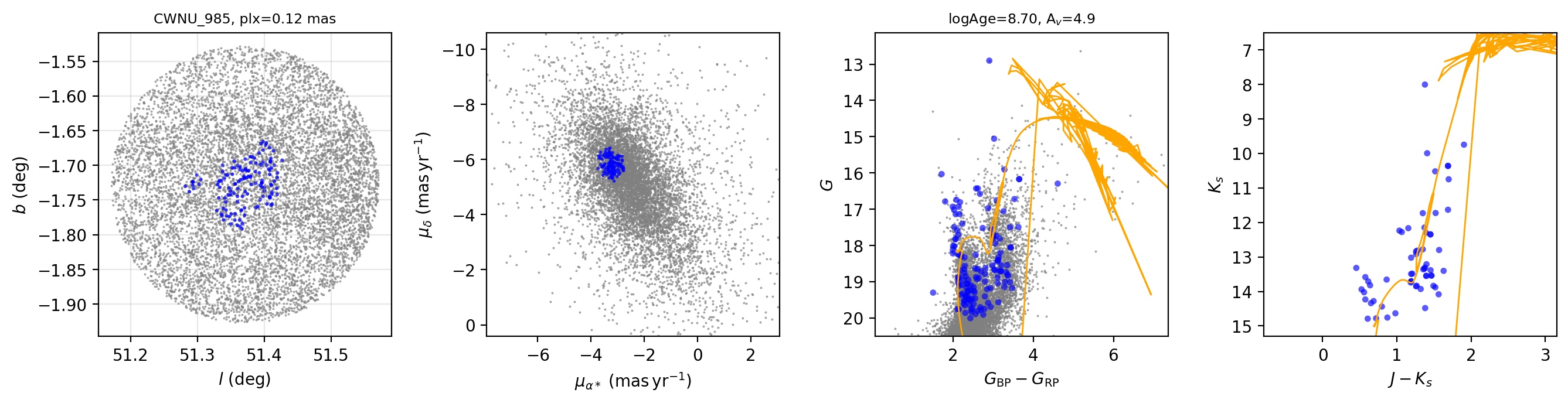}
\caption{Examples of class~2 (CWNU~791, top) and class~3 (CWNU~985, bottom). Each row shows the Galactic distribution, proper motion, Gaia CMD, and 2MASS CMD.}
\label{fig:class23}
\end{center}
\end{figure*}

The quality classes correspond to visible differences in the CMDs. As shown in Figure~\ref{fig:class23}, CWNU~791 (class~2) has a more contaminated CMD than the class~1 examples, although its center is still well defined. CWNU~985 (class~3) is larger and richer in members, but its spatial distribution appears relatively sparse. The class~2 and class~3 labels therefore reflect the difficulty of the CMD and the central concentration.

\begin{figure*}
\begin{center}
\includegraphics[width=0.62\textwidth]{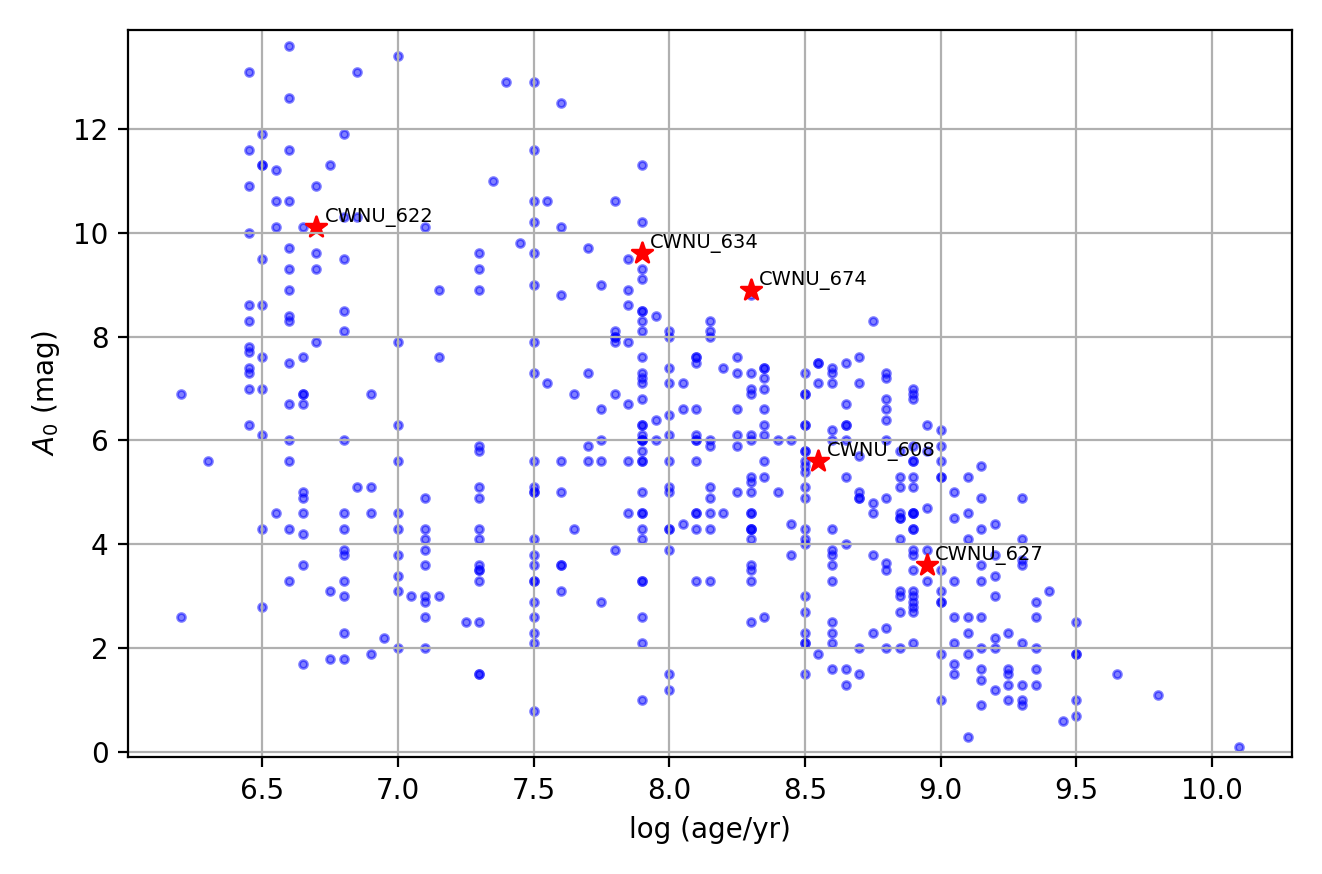}
\caption{Extinction $A_V$ against $\log$ age for the 486 objects. The five red stars are the examples shown in Figure~\ref{fig:agea0ex}.}
\label{fig:agea0}
\end{center}
\end{figure*}

\begin{figure*}
\begin{center}
\includegraphics[width=0.95\textwidth]{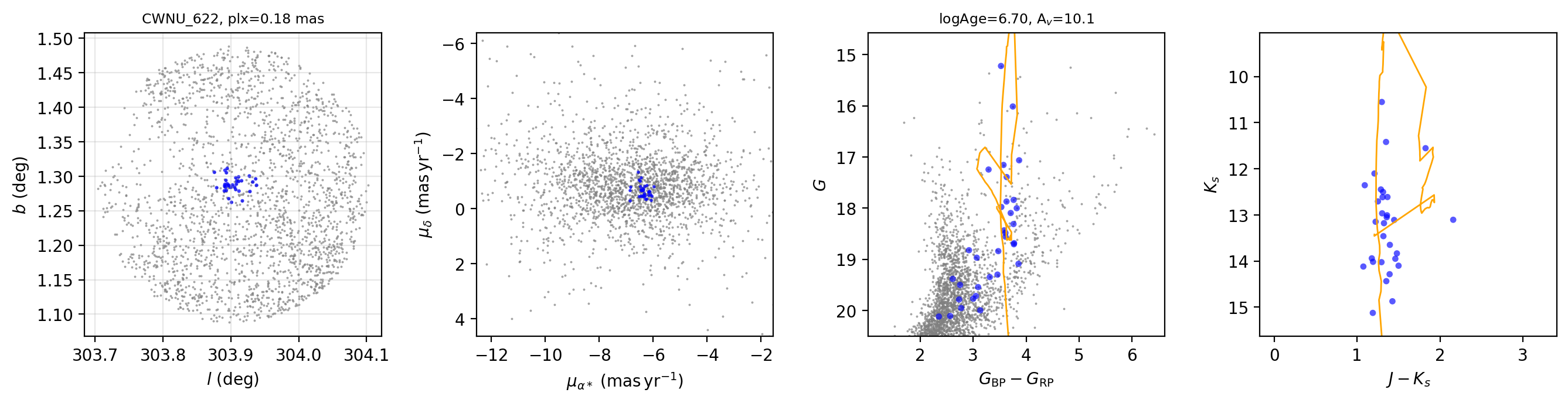}
\includegraphics[width=0.95\textwidth]{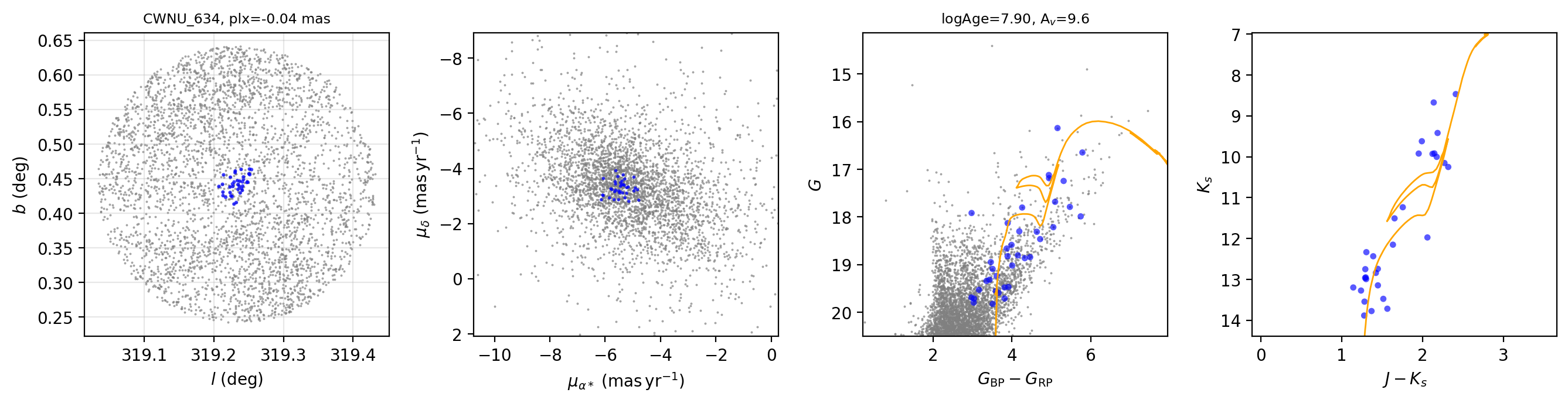}
\includegraphics[width=0.95\textwidth]{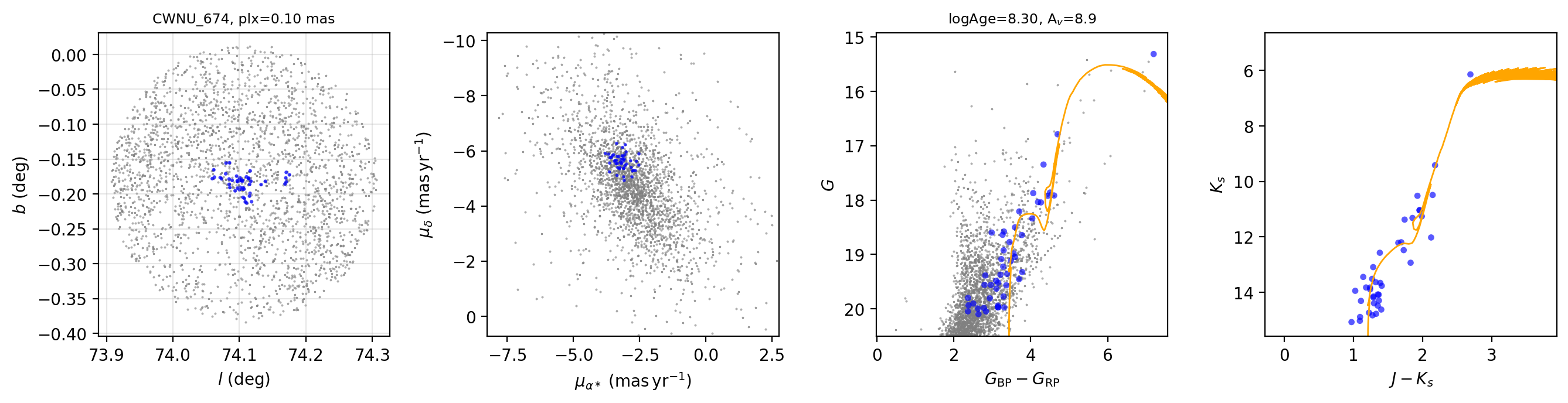}
\includegraphics[width=0.95\textwidth]{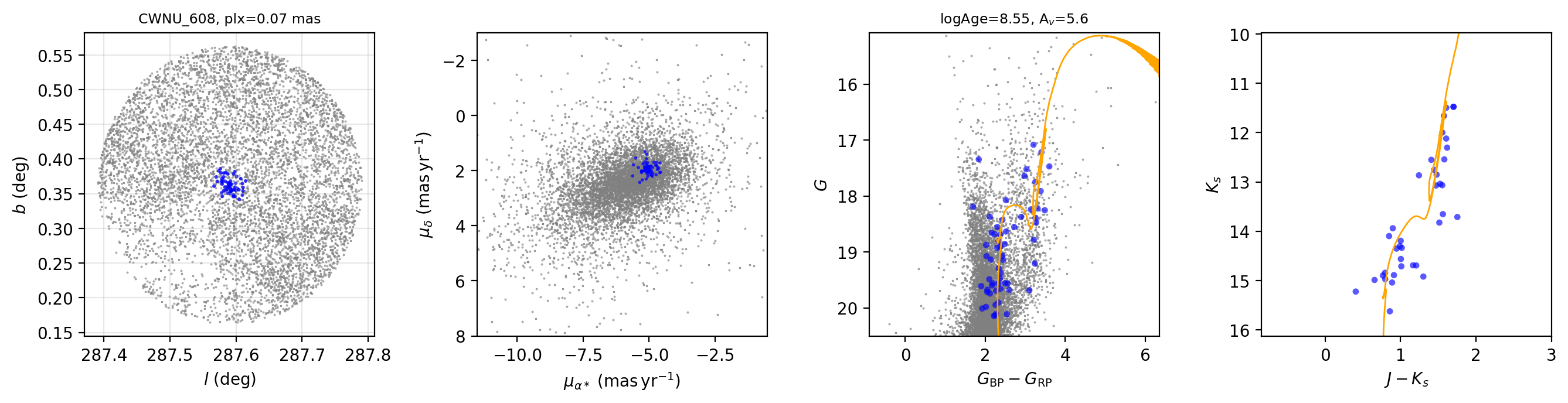}
\includegraphics[width=0.95\textwidth]{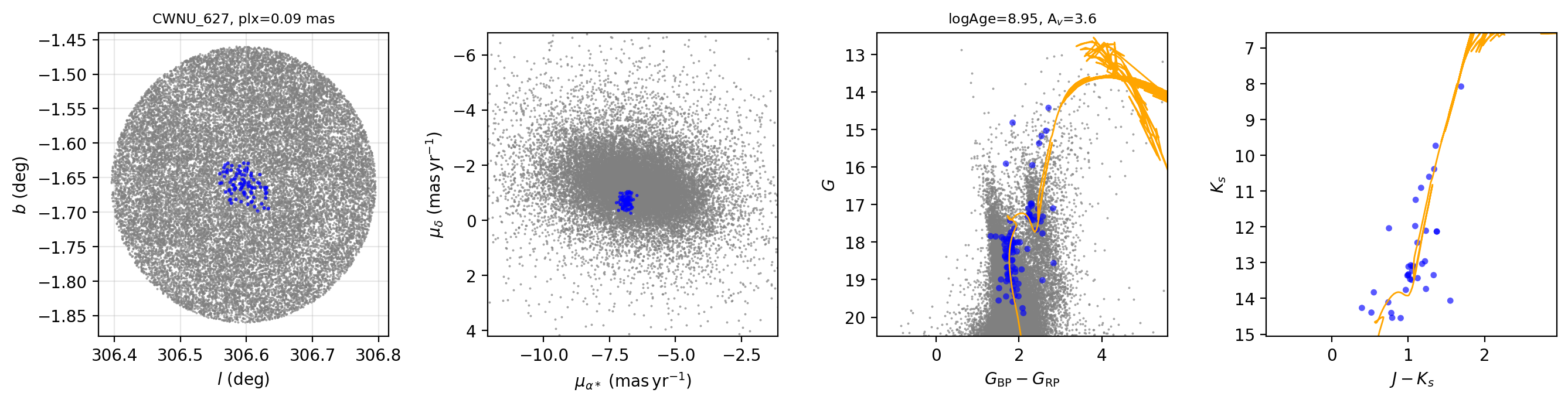}
\caption{Five clusters along the age--extinction sequence: CWNU~622, CWNU~634, CWNU~674, CWNU~608, and CWNU~627. Each row shows the Galactic distribution, proper motion, Gaia CMD, and 2MASS CMD.}
\label{fig:agea0ex}
\end{center}
\end{figure*}

The sample also shows a clear relation between age and extinction. The median $A_V$ falls from 7.4~mag at $\log$ age 6--7 to 2.5~mag at 9--10; the trend is smooth and not driven by a few objects. Figure~\ref{fig:agea0} shows the full distribution, and Figure~\ref{fig:agea0ex} shows five clusters along the sequence, from CWNU~622 (young, $A_V=10.1$~mag) to CWNU~627 (old, $A_V=3.6$~mag). The relation is expected if the older clusters in the sample are preferentially found at high latitude or in less dusty sightlines, and we return to it in Section~\ref{sec:discussion}.

\subsection{Comparison with previous catalogues}

\begin{figure*}
\begin{center}
\includegraphics[width=\textwidth]{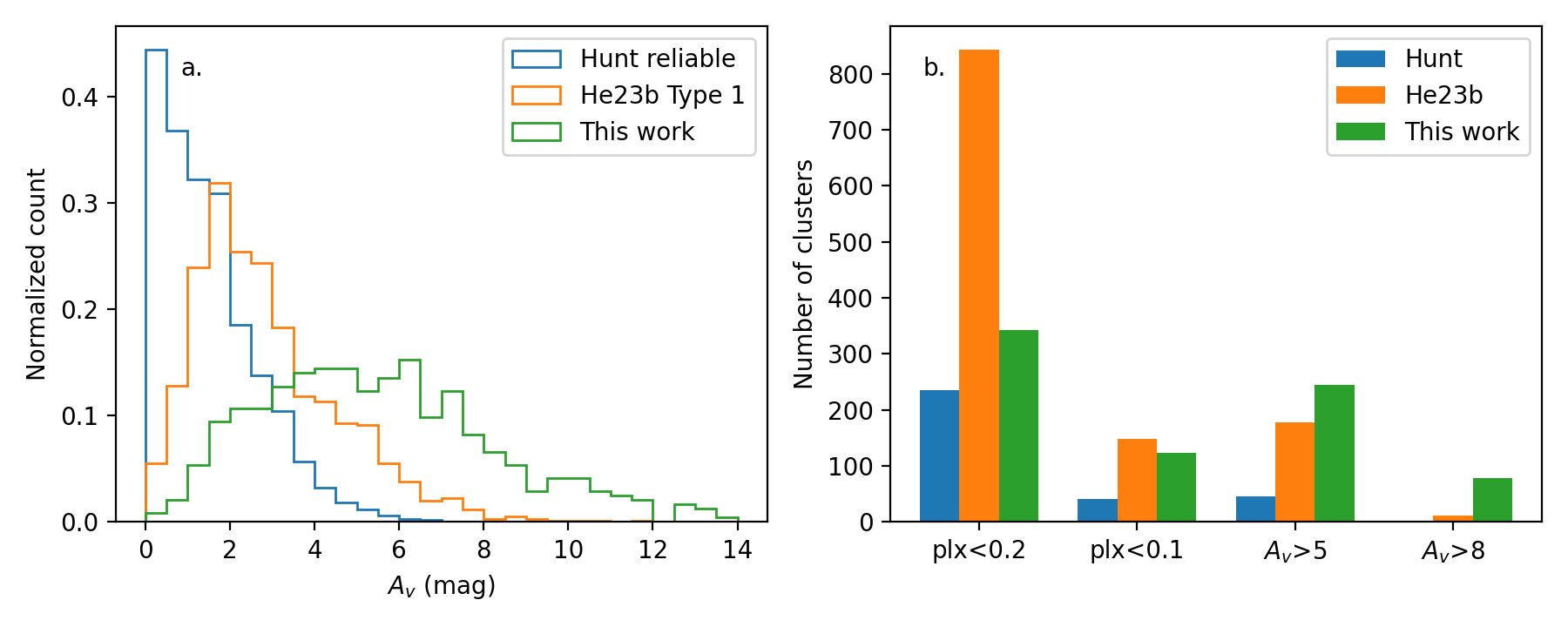}
\caption{(a) Normalized $A_V$ distributions of the Hunt reliable sample, the He23b Type~1 sample, and this work. (b) Counts at $0<\varpi<0.2$~mas, $0<\varpi<0.1$~mas, $A>5$~mag, and $A>8$~mag.}
\label{fig:comp}
\end{center}
\end{figure*}

A comparison with previous catalogues places the new sample in context. We use the reliable sample of \citet{hunt23} (4105 clusters with cluster significance CST $>5\sigma$ and median CMD class $>0.5$) and the Type~1 clusters of \citet{he23b} (1488 clusters). These samples are much larger than ours, so the comparison is not a like-for-like completeness test. We therefore show normalized $A_V$ distributions in panel (a) and absolute counts in panel (b) of Figure~\ref{fig:comp}.

The most striking difference is the high-extinction tail. Only 46 Hunt clusters and 178 He23b Type~1 clusters have $A>5$~mag, compared with 245 in our sample; at $A>8$~mag the numbers are 0, 12, and 78. The new sample therefore adds substantially to the dustiest part of the cluster census, where previous optical searches are most incomplete.

The low-parallax counts tell a different story. We have 343 objects with $0<\varpi<0.2$~mas, compared with 235 in Hunt and 843 in He23b; at $0<\varpi<0.1$~mas the numbers are 41, 148, and 124. The He23b sample already covers many distant objects, while our sample is competitive at the lowest parallaxes and much larger at high extinction. 

\section{Discussion}\label{sec:discussion}

\subsection{Completeness and contamination}

The sample is not volume-limited. Its completeness depends on distance, extinction, and stellar density. A cluster is easier to detect if it is rich, compact, and in a low-density field. At a fixed richness, it is easier to detect if the extinction is low. The class flag makes these selection effects visible: class~1 objects are the most secure, class~2 objects are plausible but contaminated, and class~3 objects are the least secure.

The main sources of contamination are field stars with proper motions similar to the cluster, unresolved binaries, and chance alignments of sparse groups. The visual inspection removes the most obvious cases, but it cannot remove all of them. A spectroscopic follow-up of a subset, especially of the class~3 objects, would be useful. The member lists also include stars with intermediate membership probabilities. The member counts should therefore be treated as lower limits for the richest objects and as upper limits for the sparsest ones.

\subsection{The warp and the extinction distribution}

The north--south displacement in Figure~\ref{fig:lb} is consistent with the Galactic warp \citep{he23warp}. The high-latitude groups lie at $l\approx60^\circ$--$110^\circ$ and $240^\circ$--$290^\circ$. In these directions the dust column above the plane is small, so clusters at large distance can still be detected. This is why the warp is seen mainly at high latitude and at large distance.

The extinction distribution also follows the large-scale dust structure. In the second and third quadrants ($90^\circ<l<270^\circ$) the extinction is on average lower (median $A_V=3.3$~mag). In the first and fourth quadrants ($l<90^\circ$ or $l>270^\circ$) the stellar density and the extinction are higher (median $A_V=6.0$~mag). The $A_V$ map in Figure~\ref{fig:lb} shows this pattern. The warp and the dust distribution therefore work together: the warp is easier to see where the dust is thin, and the inner disk is easier to sample where the clusters are bright.

\subsection{Implications for Gaia DR3 and DR4}

The main result of this paper is that Gaia DR3 still contains many clusters in the dusty disk. The 450 new objects are not a small addition. They increase the number of clusters at high extinction and low parallax by a large factor, and they fill a part of the distance--extinction plane that previous catalogues cover poorly. The CANDISC search of \citet{obasi26} covers about 1100~deg$^2$ of the inner bulge and disk and reports 40 candidates, while our Gaia search covers the disk and adds 450. The two samples are complementary.

This result has two implications. First, the completeness of the cluster census should be discussed separately for the nearby, low-extinction regime and for the distant, high-extinction regime. The former is close to complete for rich clusters; the latter is not. Recent completeness studies of the Gaia cluster census point in the same direction \citep{hunt25}. Second, the sample gives a baseline for Gaia DR4. The same clusters will be measured with better parallaxes and proper motions, and new clusters will be found in the same regions. A comparison between the DR3 and DR4 catalogues will allow the selection function to be tested directly.

\section{Summary}\label{sec:summary}

We searched for star clusters in the Galactic disk with Gaia DR3 astrometry and photometry, together with 2MASS $J$ and $K_s$ photometry. The main results are as follows.

\begin{enumerate}
\item We find 486 clusters and candidates. Of these, 450 have no literature counterpart within $r_{90}$ and are new; 36 keep their literature names.
\item The sample contains 264 class~1, 194 class~2, and 28 class~3 objects, and 24\,749 member stars. The classes are based on central concentration and CMD quality, not on extinction.
\item The median distance is 6.92~kpc, and 39\% of the objects are beyond 8~kpc. The median extinction is $A_V=5.10$~mag, and 16\% of the objects have $A_V>8$~mag.
\item The spatial distribution shows the north--south displacement of the Galactic warp. The high-latitude objects are on average farther, older, and less extincted.
\item The new sample adds many clusters at high extinction and low parallax. Gaia DR3 still has room for distant, dust-obscured clusters, and the sample gives a baseline for Gaia DR4.
\end{enumerate}

The cluster catalogue (486 objects) and the member-star catalogue (24\,749 stars) are provided as machine-readable tables with this paper. The full figure set showing the l-b and CMD views of every object will be made available with the online version.

\section*{Author Contribution}

Zhihong He conceived the methodology, and was responsible for the data processing and writing of this paper. Xiaoping Wang and Qian Cui assisted with the retrieval of the 2MASS data and the verification of the cross-match. Yangping Luo and Kun Wang reviewed the results of this paper and provided their comments.

\acknowledgements
This work was supported by National Natural Science Foundation of China (NSFC) through grants 12573023, the Sichuan Science and Technology Program Grant No. 2026YFTX0019, the “Young Data Scientists” project of the National Astronomical Data Center (NADC2023YDS-07), and the Fundamental Research Funds of China West Normal University (23kc011, 25KE031); Y.L. is supported by the NSFC12573035,12173028, and the China Space Station Telescope project: CMS-CSST-2021-A10. K.W. is supported by NSFC12373035. And this work has made use of data from the European Space Agency mission Gaia, processed by the Gaia Data Processing and Analysis Consortium (DPAC). Funding for the DPAC has been provided by national institutions, in particular the institutions participating in the Gaia Multilateral Agreement.
\endacknowledgements

\bibliography{newoc_2026}
\bibliographystyle{raa}

\label{lastpage}

\end{document}